\documentclass[12pt, doublespacing]{article} 

\usepackage[latin1]{inputenc}
\usepackage{graphicx, geometry,  amsmath, amsthm, amssymb,    amsfonts,  subfigure, amstext}
\usepackage{wrapfig, amsopn, latexsym,  epsfig, multirow, multicol, verbatim}
\usepackage{fancyhdr,  bm, caption}
\usepackage[semicolon]{natbib}
\usepackage{booktabs,caption, comment}
\usepackage[flushleft]{threeparttable}

\allowdisplaybreaks

\newcommand{\blind}{0}

\usepackage{natbib} 
\usepackage[plainpages=false,pdfpagelabels,hypertexnames,linktocpage, colorlinks, citecolor=red,linkcolor=blue]{hyperref}

\begin{document}
 
\setlength{\parindent}{0pt}
\setlength{\parskip}{5pt}

\def\spacingset#1{\renewcommand{\baselinestretch}%
{#1}\small\normalsize} \spacingset{1}


\if0\blind
{
  \title{\bf Inverse stable prior for exponential models}
  \author{Dexter Cahoy \\ 
    Department of Mathematics and Statistics \\  University of Houston-Downtown \\  Houston, TX 77002\\ 
\texttt{cahoyd@uhd.edu}\\
    \\ 
    Joseph Sedransk\\
    Joint Program in Survey Methodology \\  University of Maryland \\ College Park, MD 20742\\
\texttt{jxs123@case.edu} }
  \maketitle
} \fi

\if1\blind
{
  \bigskip
  \bigskip
  \bigskip
  \begin{center}
    {\LARGE\bf Title}
\end{center}
  \medskip
} \fi

\bigskip
\begin{abstract}

We consider a class of non-conjugate priors as a mixing family of distributions  for a  parameter (e.g.,  Poisson or gamma rate, inverse scale or precision of an inverse-gamma, inverse variance of a normal distribution)  of an exponential subclass of discrete and continuous data distributions.    The prior class is proper,  nonzero at the origin (unlike the gamma and inverted beta priors with shape parameter less than one and Jeffreys prior for a Poisson rate), and is easy to generate random numbers from.  The  prior class  also  provides  flexibility  in capturing a wide array of  prior beliefs (right-skewed and  left-skewed)  as modulated by a bounded  parameter $\alpha \in (0, 1).$   The resulting posterior  family in the single-parameter case can be expressed in closed-form and is  proper,  making calibration unnecessary.  The mixing induced by the inverse stable family results to a marginal prior distribution in the form of a generalized Mittag-Leffler function, which covers a broad array of distributional shapes.    We derive closed-form expressions of some properties  like the moment generating function and   moments.   We propose algorithms to generate samples from the posterior distribution and   calculate the Bayes estimators  for real data analysis.  We  formulate the predictive prior and posterior distributions. We test the  proposed Bayes estimators   using Monte Carlo simulations.   The extension to  hierarchical modeling  and inverse variance components models  is straightforward. We explore  the global shrinkage model in some detail to show the potential value of the inverse stable prior.    We show that the inverse stable prior    has some better properties than the inverted beta prior corresponding to the half-Cauchy prior (commonly recommended for adoption in such cases). We illustrate the methodology using a real data set, introduce a hyperprior density for the hyperparameters, and extend the model to  a heavy-tailed  distribution.

\end{abstract}

\noindent%
{\it Keywords:}   rate, inverse scale, inverse variance, precision, Jeffreys,       Mittag-Leffler 
\vfill

\section{Introduction}

We illustrate some motivations  of this paper using  common practical applications of the widely used  conjugate {\tt gamma} ($\nu_0$, $\lambda_0$) prior density 
\begin{equation}
p(\lambda \; | \; \nu_0, \lambda_0 ) \; =\; \frac{\lambda_0^{\nu_0}}{\Gamma ( \nu_0) } \lambda^{\nu_0-1} e^{-\lambda_0 \lambda},  \qquad \qquad  \lambda >0,  \nu_0 >0,\;  \lambda_0 >0. \label{gammaprior}
\end{equation}
If $\nu_0<1$ and as $\lambda \to 0$, the limiting density fails to  exist.  Therefore,  an infinite mass  at the origin introduces bias toward values near the origin.  

{\it Application 1.} If a discrete dataset $\bm{X} =\left( X_1, X_2, \ldots, X_n \right)$ is a random sample from {\tt Poisson} ($\lambda$) with probability mass function $f(x | \lambda)= \frac{\lambda^x e^{-\lambda }}{x!}$ then we obtain the {\tt Poisson}($\lambda$)-{\tt gamma} ($\nu_0$, $\lambda_0$) model  where  the posterior density  is   proportional to 
\begin{equation}
  \lambda^{\sum_{j=1}^n x_j -(1 -\nu_0)} e^{-(n+\lambda_0) \lambda}. 
\end{equation}
As $\lambda \to 0$   and whenever $\sum_{j=1}^n x_j <(1-\nu_0),$ the posterior diverges (with infinite limit).  This  posits a limitation  on   inference for the {\tt Poisson-gamma} model.  A true zero rate (or infinite variance) cannot be ignored but  an infinite mass  cannot be assigned only  at/near the origin  as well.  Note that observing $\sum_{j=1}^n x_j  =0$ in practice,  for instance, is not uncommon especially if a relatively small dataset is sampled from a zero-inflated  population or from  a Poisson distribution with mean $\lambda$ close to zero.  

Furthermore, the {\tt Jeffreys} prior \citep[see also][]{dan99}  $p(\lambda) \sim  \lambda^{-1/2}$ for the Poisson distribution also tends to  infinity as $\lambda \to 0.$     The corresponding posterior kernel is 
\begin{equation}
  \lambda^{\sum_{j=1}^n x_j -1/2} \; e^{-n \lambda},
\end{equation}
which again suffers from the same drawback as $\lambda \to 0$ whenever $\sum_{j=1}^n x_j =0.$

{\it Application 2.}  If  $\bm{X}$ is continuous and is a random sample from  {\tt gamma} ($\nu,  \lambda$)  with probability density function $ f(x | \lambda, \nu)= \frac{\lambda^{\nu}}{ \Gamma (\nu)}\;  x^{\nu-1} e^{-\lambda x}$  then we have  the {\tt gamma} ($\nu,  \lambda$)-{\tt gamma} ($\nu_0,  \lambda_0$) model. The posterior density kernel is 
\begin{equation}
  \lambda^{\nu_0 + n \nu-1 } e^{-[\lambda_0 + \sum_{j=1}^n x_j] \lambda}. 
\end{equation}
If both shape parameters ($\nu_0$ and $\nu$) above are relatively small and such that $\nu_0 + n \nu <1,$  then the limit ceases to  exist  also  as $\lambda \to 0.$ 

{\it  Application 3.}   If  the sample data $\bm{X}$  come from  {\tt inverse-gamma} ($\nu,  \lambda$)  with probability density function $f(x | \lambda, \nu)= \frac{\lambda^\nu}{ \Gamma (\nu)}\;  x^{-\nu-1} e^{-\lambda/x}$ then we obtain the  {\tt  inverse-gamma} ($\nu , \lambda$)-{\tt gamma} ($\nu_0,  \lambda_0$)  model for the inverse scale  parameter  $\lambda,$  which yields another gamma posterior  kernel
\begin{equation}
  \lambda^{\nu_0 + n \nu-1 } e^{-[\lambda_0 + \sum_{j=1}^n 1/ x_j] \lambda}. 
\end{equation}
If the same  conditions  in the second application above  are satisfied then the posterior suffers from the same drawback as well.

{\it  Application 4.}     Consider 

\begin{equation}
X_j \; | \;  \lambda_j \stackrel{d}{=}  N(\lambda_j, 1), \quad \Lambda_j \; | \; \delta^2 \stackrel{d}{=}  N (0,  \delta^2),   \label{ib}
\end{equation}
with the half-Cauchy prior analogue for $\delta^2$, the inverted beta    prior \citep{gel06, pas12b}   
\begin{equation}
\delta^2 \;  \sim \; \left(\delta^2 \right)^{-1/2} \left( 1+ \delta^2\right)^{-1}. \label{ib2}
\end{equation} 
The prior explodes at the origin suffering from the same pitfall as seen above.  Simple algebra yields the posterior
\begin{equation}
\delta^2 \;  |   \; data \;\sim \;  \left(\delta^2 \right)^{-1/2} \left( 1+ \delta^2\right)^{-\frac{n+2}{2}} \exp \left( - || \bm{x}||_2^2/[ 2 (1 + \delta^2)]\right)  \label{ib3}
\end{equation} 
 which is again explosive at zero.  Here,  $|| \cdot ||_2$ is the $L_2$-norm.  For applications, such as those given above,  finding priors  whose corresponding posterior distributions have better posterior properties near the origin is important. This is a motivation of this paper together with providing informative and proper  priors  for the rate parameter of frequently used exponential models. 
 
 The rest of the paper is organized as follows. The inverse stable density is introduced in Section 2.  The single parameter formulation is presented in Section 3.   The  computational algorithms  are  given in Section 4. Predictive distributions  are discussed  in Section 5.  The extensions to hierarchical settings (particularly the inverse stable
analogue of the global shrinkage model based on an inverse scale
mixture of normals) are explored in more depth in Section 6.  Illustrations using real data  are in Section 7.     The extension  to  heavy-tailed  prior specification  and  some concluding remarks  are presented in Section 8.

\section{Inverse stable density}

The  inverse stable (IS) density has been increasingly becoming popular in several areas of study particularly in physics and mathematics.    It  is a probability model   for  time-fractional  differential equations, which  leads to closed-form solutions \citep{psw05, mmp10, mnv11, mas13, iks17}.  It is also  used as a subordinator (as the operational time rather than the physical time) for time-fractional diffusions and for Poisson-type processes \citep{bao10}.  

The inverse stable function is related to the strict $\alpha^+$-stable density (see all  references above) as follows: If 
$
\Theta \stackrel{d}{=} IS_{\alpha, \rho} ( \theta)
$
then 
\begin{equation}
\Theta\; \stackrel{d}{=} \; \frac{\rho^{1/\alpha} \; \theta^{-1-1/\alpha}}{\alpha} \;  g_\alpha\left(\;   \rho^{1/\alpha}\; \theta^{-1/\alpha} \; \right), \qquad    \theta >0,  \label{stabform}
\end{equation}
where the Laplace transform of the $\alpha^+$-stable density $g_\alpha(s)$ is $\phi_S (\beta) =\mathbf{E}( e^{-\beta s} )= \int_{\mathbb{R}^+} e^{-\beta s} g_\alpha(s) ds= e^{-\beta^\alpha}, \alpha \in (0, 1).$  Note that the Airy ($\alpha=1/3$) and  half-normal ($\alpha=1/2$)  distributions are special cases.  As $\alpha \to 1^-$, the distribution becomes degenerate, i.e., 
\begin{equation}
  IS_{\alpha, \rho} ( \theta) \longrightarrow \delta (\theta -\rho).
\end{equation}
The prior family (indexed by $\alpha$) does not vanish at $\theta = 0^+$, i.e,  $IS_{\alpha, \rho} ( \theta) \to \rho^{-1}/\Gamma(1-\alpha), \theta \to  0^+$ (unlike the gamma prior with shape parameter less than one).   The  large-sample behavior \citep{mmp10,mas13}  of $IS_{\alpha,1}(\theta)$ can be simply expressed in terms of $\theta/\alpha$  as 

\begin{equation}
 IS_{\alpha, 1} \left( \theta / \alpha \right) \sim  c_1(\alpha) \theta^{ (\alpha-1/2 )/(1-\alpha)} \exp [ - c_2(\alpha) \theta^{1/(1-\alpha)} ],
\end{equation}

where  $c_1(\alpha)= (2 \pi \alpha (1-\alpha) )^{-1/2} >0,$ and $c_2(\alpha) =(1-\alpha)/\alpha >0.$    Moreover, 
\begin{equation}
  \mathbf{E} \Theta^k = \frac{\rho^k   \Gamma (1 + k)}{\Gamma (1 + \alpha k)}, \qquad \phi_\Theta (\beta)= E_\alpha (-\beta \rho),  \qquad \phi_\rho (\beta)= \beta^{1-1/\alpha} e^{- \theta \beta}, \label{2.1}
\end{equation}
where $k>-1, 
\;  \text{and}\;    E_\alpha (u )=  \sum\limits_{j=1}^\infty \frac{u^j}{j! \;  \Gamma (1+ \alpha j )} \label{MitLef}
$ is the Mittag-Leffler function \citep{bing71}.   Random variates from  (\ref{stabform})   can be generated using the structural representation  
\begin{equation}
\Theta \stackrel{d}{=}\rho   S^{-\alpha}, \; \qquad    \;   S \stackrel{d}{=}  g_\alpha(s).
\end{equation}
The random variable $S$ can be generated using the following formula \citep{kan75, cms76}:
 
\begin{equation}
S \stackrel{d}{=}  \frac{\sin(\alpha \pi U_1)[ \sin((1-\alpha)\pi
U_2)]^{1/\alpha-1}}{[\sin (\pi U_2)]^{1/\alpha}|\ln
U_1|^{1/\alpha-1}}, 
\end{equation}
where $U_1$ and $U_2$   are independently and uniformly distributed in $[0,1]$.

\begin{figure}[h!t!b!p!]
     \centering
                         \includegraphics[height=2.5in, width=5in]{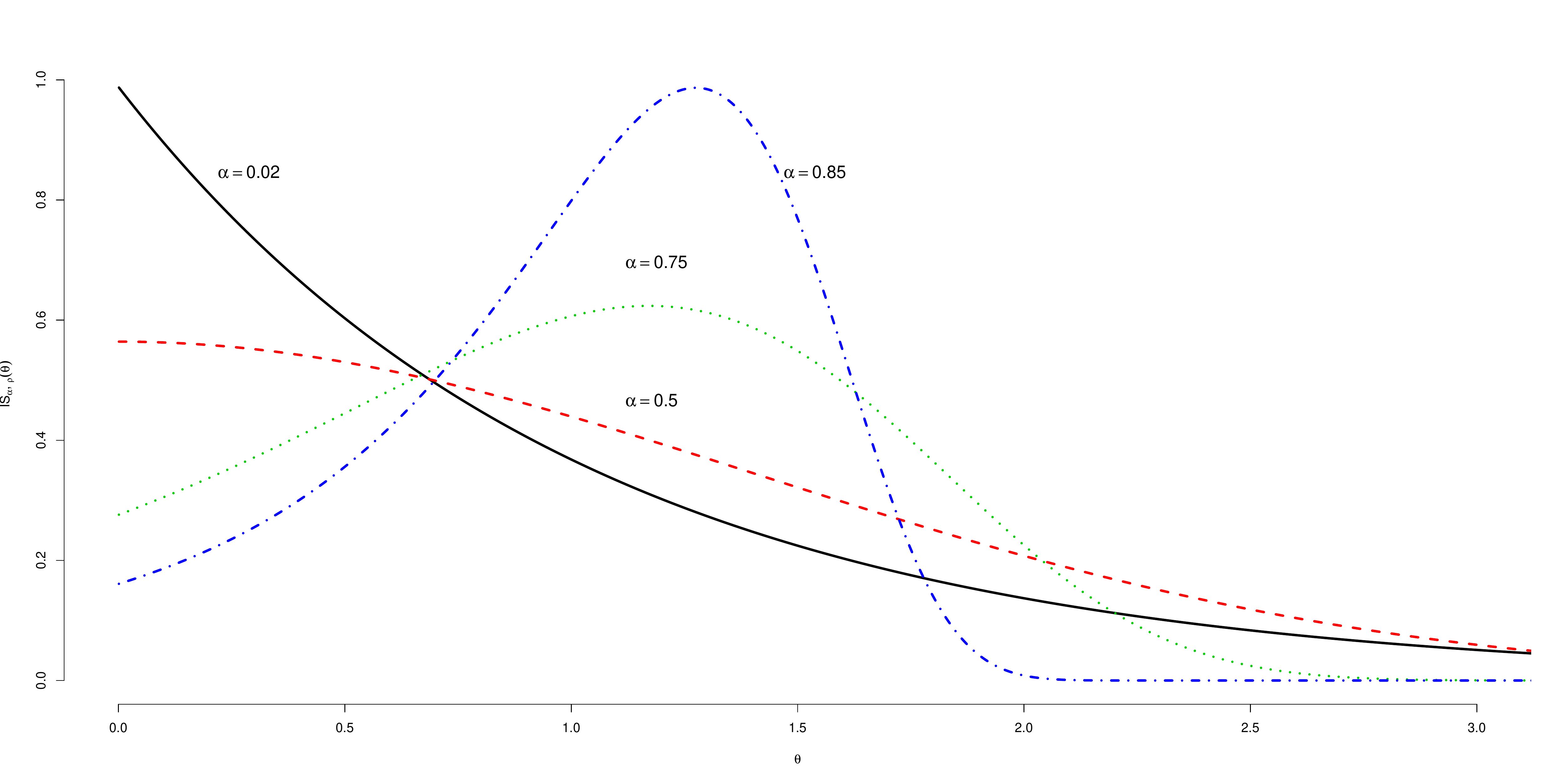}
       \renewcommand\abovecaptionskip{0pt}
  \caption{Inverse stable prior  densities  with $ \rho=1$.}
\label{f1}
\end{figure}

 Figure \ref{f1} below reveals some members of the prior family as a function of the hyperparameter $\alpha.$ 
Apparently,  $\alpha$ controls the shape of $IS_{\alpha, \rho} ( \theta).$  For smaller values 
of $\alpha$,  $IS_{\alpha, \rho} ( \theta)$  assigns finite masses  near the origin and  becomes bounded as $\alpha \to 0^+.$   The distributional shapes include right-skewed and  left-skewed densities.  Clearly,  the class of priors allows  considerable  flexibility in capturing prior beliefs than the widely used gamma density.

\section{Single parameter formulation}

\emph{Proposition 1.}   Let $ \bm{X} =\left( X_1, X_2, \ldots, X_n \right) $ be a random sample  from a population  belonging to an exponential family of distributions  with the  likelihood kernel

\begin{equation}
L(\theta | \bm{x}) \; \sim \;     e^{-a \cdot \theta \; + b \;  \cdot \log (\theta) },  
\end{equation}

 where  $a=a(\bm{x}) ,  b=b(\bm{x}) \in \mathbb{R}^+ \cup \{0\}, \theta =  | \theta(\epsilon)|,  \theta(\epsilon)   \in \mathbb{R} \backslash \{ 0\}$. Using  the  non-conjugate inverse stable density
$p(\theta)=IS_{\alpha, \rho } ( \theta)$ as a  prior   for $\Theta$  yields the the proper or normalized posterior distribution
\begin{equation}
p(\theta |  \bm{x}  ) \;  =  \;  \frac{e^{-a \cdot \theta \; + \; b \cdot \log (\theta)}  \;  IS_{\alpha, \rho} \left(  \theta   \right) }{\Gamma ( b+1)  \; \rho^{b} \; E_{\alpha,\alpha b + 1}^{b+1} (-a \rho )  }, \label{post} 
\end{equation}
where  $\alpha$ and $\rho$ are the hyperparameters, 

\begin{equation}
			E_{\eta,\nu}^\tau( w) = \sum_{j=0}^\infty \frac{(\tau)_j}{j!\Gamma(\eta j+\nu)} \;  w^j 
\end{equation}
is the generalized Mittag--Leffler function \citep{pra71},    $ \eta,\nu,\tau, w  \in \mathbb{C};  \Re(\eta),\Re(\nu),\Re(\tau) >0$,   $(\tau)_j = \Gamma ( \tau + j  )/  \Gamma ( \tau )  $ is the Pochammer symbol, and $(\tau)_0=1.$

\emph{Proof.} See Appendix. 

For the {\tt Poisson-gamma} model in Section 1, as $b=\sum_{i=0}^n x_i \to 0 $ and   $\theta \to 0$, the posterior $p(\theta|\bm{x} ) \to  \rho^{-1}/\left[ \Gamma (1-\alpha)  E_{\alpha, 1}^{1} (-a \rho )   \right]$, i.e.,  the limit exists.   For both the {\tt gamma-gamma}  and {\tt inv-gamma-gamma} models above,  as $b= n \nu  \to 0$, i.e.,  as  $\nu \to 0,$  the posterior $p(\theta|\bm{x} ) \to  \rho^{-1}/\left[ \Gamma (1-\alpha)  E_{\alpha, 1}^{1} (-a \rho )   \right]$ as well as $\theta \to 0$. 

\emph{Corollary 1.}   The marginal density of the data given hyperparameters $\alpha$ and $\rho$  is  
\begin{equation}
p(\bm{x} | \alpha, \rho) \;  =  \;   \Gamma ( b+1)  \; \rho^{b}\;  E_{\alpha,\alpha b + 1}^{b+1} (-a \rho ). \label{gml2}
\end{equation}
 
\emph{Proof.}  It is trivial and is omitted. \qed 

The  generalized Mittag-Leffler function in (\ref{gml2}) above is absolutely convergent for all $a$ \citep[see][]{sap97}. If  $b=0$ then we obtain the Mittag-Leffler function.

\emph{Corollary 2.}   The $k$th posterior moment  is 

\begin{equation}
\mathbf{E}\left(  \Theta ^k \big \vert \bm{x} \right)\;  =  \;  \frac{ \Gamma (k+b+1) \; \rho^{k} \; E_{\alpha,\alpha (k+b) + 1}^{k+ b+1} (-a \rho ) }{\Gamma ( b+1)  \;  E_{\alpha,\alpha b + 1}^{b+1} (-a \rho )  }.
\end{equation}

\emph{Proof.}  It is trivial and thus is not provided. \qed 

When $k=0$, the formula in (\ref{post}) is checked as a valid probability density function.  


\emph{Corollary 3.}    The moment generating function is straightforward to calculate as
\begin{equation}
\mathbf{E} \left( \exp \left( \beta \Theta  \right)  \vert \bm{x} \right) = \frac{ E_{\alpha,\alpha b + 1}^{b+1} (-(a-\beta) \rho)  }{ E_{\alpha,\alpha b + 1}^{b+1} (-a \rho) }.
\end{equation}

\emph{Proof.}  The proof follows from the property of the inverse stable distribution and is trivial.\qed

\emph{Corollary 4.}  As $\alpha \longrightarrow 1^-$,
\begin{equation}
p(\theta |  \bm{x}  ) \;   \longrightarrow   \;  \frac{e^{-a \cdot \theta}  \;  \theta^b \; \delta \left(  \theta  -  \rho \right) }{e^{- a \rho}  \; \rho^b}  \label{post1} 
\end{equation}
because  $ E_{1^-,  (1^-) b + 1}^{b+1} (-a \rho)   = e^{ -a\rho}/\Gamma (b+1) $.

\emph{Proof.}  The proof follows from the property of the inverse stable density and is omitted. \qed 

The preceding corollary suggests that the Bayes estimator  $\widehat{\theta}_{Bayes} \;  \big|_{\alpha \to  1^-}\; \to \;\rho$.  It also indicates how  $\alpha$ controls the shape and/or variability of the posterior distribution from a  non-degenerate family $\left(\alpha = 0^+\right)$ to a  degenerate one $ \left(\alpha = 1^-\right)$.  This shows the relevance of $\rho$  as  $\widehat{\rho} \longrightarrow  \theta$  when $\alpha = 1^-$ and as $n$ goes large. 



Table \ref{t1} below shows some likelihood models with  their corresponding parameterizations following the above  formulation and notation.  The inverse stable density can also be a prior for  a rate or an inverse scale parameter of   two-parameter models like the inverse-Weibull,   lognormal  and  normal-inverse densities conditional on  the other parameter \citep[see e.g. ][]{mil80,kun08}.

\begin{table}[h!t!b!p!]
\caption{\emph{Probability models with corresponding parameterizations.}} 
\begin{tabular*}{6.8in}{@{\extracolsep{\fill}}c|c|c|c}
 \hline
 \multirow{2}{*}{Data model} &  \multirow{2}{*}{Parameter $\theta$}   &  \multirow{2}{*}{$a$}  &  \multirow{2}{*}{$b$}  \\ 
& & & \\
\hline \hline
\multirow{2}{*}{Poisson: $\lambda^x e^{-\lambda}/x!$} & \multirow{2}{*}{$\lambda$} & \multirow{2}{*}{$n$} & \multirow{2}{*}{$\sum x_i$} \\
& & & \\
\hline
 \multirow{2}{*}{Rayleigh: $ \frac{x}{\sigma^2}  \;  e^{-x^2/(2\sigma^2)}$} & \multirow{2}{*}{$1/\sigma^2$ } & \multirow{2}{*}{$\sum x_i^2/2$} & \multirow{2}{*}{$n$} \\
& & & \\
\hline
\multirow{2}{*}{Half-normal: $ \frac{2\sigma}{\pi}\;  e^{-(x^2\sigma^2)/\pi}$} &  \multirow{2}{*}{$\sigma^2$ } & \multirow{2}{*}{$\sum x_i^2/\pi$} & \multirow{2}{*}{$n/2   $} \\
& & & \\
\hline
 \multirow{2}{*}{Inverse Rayleigh: $ (2 \varepsilon^2 / x^3) \;  e^{-\varepsilon^2/x^2}$} & \multirow{2}{*}{$\varepsilon^2$ } & \multirow{2}{*}{$\sum 1/x_i^2$} & \multirow{2}{*}{$n$} \\
& & & \\
\hline
 \multirow{2}{*}{Exponential: $\lambda e^{-\lambda x}$} & \multirow{2}{*}{$\lambda$ } & \multirow{2}{*}{$\sum x_i$} & \multirow{2}{*}{$n$}  \\
& & & \\
\hline
 \multirow{2}{*}{Laplace: $ (\lambda /2) \exp ( -\lambda |x| )$} & \multirow{2}{*}{$\lambda$ } & \multirow{2}{*}{$\sum |x_i|$} & \multirow{2}{*}{$n$}  \\
& & & \\
\hline
\multirow{2}{*}{Inverse Exponential: $\lambda x^{-2} e^{-\lambda/x}$} & \multirow{2}{*}{$\lambda$} & \multirow{2}{*}{$\sum 1/ x_i$} & \multirow{2}{*}{$n$} \\
& & & \\
\hline
 \multirow{2}{*}{Skew-Logistic: $ \lambda \left(1+e^{-x} \right)^{-\lambda-1} e^{-x}$} & \multirow{2}{*}{$\lambda$ } &  \multirow{2}{*}{$\sum \log \left(1+ e^{-x_i} \right)$} & \multirow{2}{*}{$n$} \\
 & & & \\
\hline
 \multirow{2}{*}{Gamma: $ \frac{\varepsilon^\sigma}{\Gamma (\sigma)}  x^{\sigma-1}  e^{-\varepsilon x}$} & \multirow{2}{*}{$\varepsilon, \sigma$-known } & \multirow{2}{*}{$\sum x_i$} & \multirow{2}{*}{$n \sigma $} \\
& & & \\
\hline
 \multirow{2}{*}{Weibull: $ \frac{\sigma}{\varepsilon} \left(\frac{x}{\varepsilon} \right)^{\sigma-1}  e^{-(x/\varepsilon)^\sigma}$} & \multirow{2}{*}{$1/\varepsilon^\sigma, \sigma$-known } & \multirow{2}{*}{$\sum x_i^\sigma$} & \multirow{2}{*}{$n$} \\
& & & \\
\hline
\multirow{2}{*}{Normal: $ \frac{ e^{-(x-\mu)^2/(2\sigma^2)}}{\sqrt{2 \pi \sigma^2}}$} &  \multirow{2}{*}{$1/\sigma^2, \mu$-known } & \multirow{2}{*}{$\sum (x_i-\mu)^2/2$} & \multirow{2}{*}{$n/2   $} \\
& & & \\
\hline
 \multirow{2}{*}{Generalized  Exponential: $ \varepsilon \sigma \left(1-e^{-\sigma x} \right)^{\varepsilon-1} e^{-\sigma x}$} & \multirow{2}{*}{$\varepsilon, \sigma$-known } & \multirow{2}{*}{$-\sum \log \left(1- e^{- \sigma x_i} \right)$} & \multirow{2}{*}{$n $} \\
& & & \\
\hline
 \end{tabular*}
  \label{t1}
\end{table}

\section{Estimation algorithms}

\subsection{Monte Carlo integration}

From  (\ref{res1}) of Appendix A,  we have
\begin{equation}
E_{\alpha,\alpha \omega + 1}^{\omega+1} (- a \rho )\; = \; \frac{1}{\rho^{  \omega}\Gamma ( \omega +1) }\int_{\mathbb{R}^+}e^{-a    y} y^\omega \cdot IS_{\alpha,\rho} (y) dy  \; \approx \; \frac{\sum_{j=1}^J e^{-a     Y_j} Y_j^{\omega} }{J \;\rho^{  \omega} \; \Gamma (\omega+1)},      \label{mcif}
\end{equation}
where   $Y_j's \; \stackrel{ind}{=} \rho S^{-\alpha}.$  Thus, the $k$th moment can be estimated  using the following algorithm.

\textbf{Algorithm 1}

\textbf{Step 1.} Generate   $Y_1, Y_2, \ldots,Y_J \stackrel{ind}{=}    \rho  S^{-\alpha}.$ 

\textbf{Step 2.} Let
\begin{equation}
V_{\alpha, \rho, k} =  \frac{\sum_{j=1}^J e^{-a    Y_j} Y_j^{b+k} }{J \;\rho^{ (b+k)} \; \Gamma (k+b+1)}.
\end{equation}

\textbf{Step 3.} Set
\begin{equation}
\widehat{\mathbf{E} \left(  \Theta^k \big \vert \bm{x} \right)  } =  \frac{  \Gamma (b+k+1) \; \rho^{  k} \; V_{\alpha, \rho,k} }{ \Gamma (b+1)\; V_{\alpha, \rho,  0} }. 
\end{equation}

In particular, the Bayes estimator of $\theta$ with respect to the squared error loss criterion can be directly computed     as

\begin{equation}
\widehat{\theta}_1=  \frac{ (b+1) \;  \rho  \; V_{\alpha, \rho, 1} }{ V_{\alpha, \rho,  0} } 
\end{equation}
while the variance can be estimated as 

\begin{equation}
\widehat{\mathbf{Var}  \left(  \Theta \big \vert \bm{x} \right) } \; = \;  \frac{(b+1) \rho^{2 \alpha}}{V_{\alpha, \rho,  0} }\left[  (b+2)    V_{\alpha, \rho, 2}  -  \frac{ (b+1) \left(  V_{\alpha, \rho, 1}  \right)^2}{ V_{\alpha, \rho,  0} }\right]. 
\end{equation}


\subsection{Posterior simulation}

Let the target density be $p(\theta |  \bm{x} ) $ and candidate density be $IS_{\alpha, \rho}(\theta), \alpha$ and $\rho$ are both known. Then
\begin{equation}
\frac{p(\theta |  \bm{x} ) }{IS_{\alpha, \rho}(\theta)}  \;  =  \;  \frac{e^{-a \cdot \theta}  \;  \theta^b }{\Gamma ( b+1)  \; \rho^{ b} \; E_{\alpha,\alpha b + 1}^{b+1} (-a \rho )  }.
 \end{equation}
The  preceding expression is maximized at $\theta=b/a$ and 
\begin{equation}
 \frac{e^{-a \cdot \theta}  \;  \theta^b }{\Gamma ( b+1)  \; \rho^{  b} \; E_{\alpha,\alpha b + 1}^{b+1} (-a \rho )   }  \; \;  < \;\;   \frac{e^{-b}  \;  (b/a)^b }{\Gamma ( b+1)  \; \rho^{ b} \; E_{\alpha,\alpha b + 1}^{b+1} (-a \rho )    } \;\;=\;\; Q.
\end{equation}
Below is an accept-reject algorithm   for generating observations from the posterior distribution (\ref{post}). 


\textbf{Algorithm 2}

\textbf{Step 1.} Generate $U  \stackrel{d}{=}  U(0,1)$ and $Y \stackrel{d}{=}  \rho \cdot  S^{-\alpha}.$ 

\textbf{Step 2.} Accept $\Theta=Y$ if  

\begin{equation}
u< \frac{p(y |  \bm{x}  ) }{Q \cdot IS_{\alpha, \rho}(y)}   \; = \;  \frac{e^{-a \cdot y}  \;  y^b }{e^{-b}  \;  (b/a)^b }\;=\; e^{b - ay }  \cdot  \left(ay/b \right) ^b
\end{equation}

or equivalently,
$ \log(u) < \log \left[  \frac{p(y|  \bm{x}  ) }{Q  \cdot IS_{\alpha, \rho}(y) } \right]  =b\; \left[\; \log(ay/b) +1 \; \right] -ay;$  otherwise, go back to step 1.

\textbf{Step 3.} Repeat steps $1-2, B$ times and set $ \widehat{\theta}_2 = \overline{\Theta}$.

 As $\alpha \to 1$, $Q=\frac{e^{-b} \;  (b/a)^b  \;  }{\rho^b\; e^{-a\rho}} =1$ if and only if  $b=a$ and $\rho =1.$  Thus, we expect high acceptance rates (close to $100\%$) of Algorithm 2  if the conditions ($\alpha \approx 1$,  $b = a$ and $\rho =1$) are close to being satisfied. The same acceptance rates are to be observed  for $\alpha \approx 1$,  $b \neq a$ but $\rho$ is the solution of    $e^{-b} \;  (b/a)^b  \;  - \rho^b\; e^{-a\rho}=0.$  

We generated 500 observations from the different  posterior distributions  in 1000 replications based on observed data sizes $n=15, 30, 60$   to test  both  algorithms above.    We calculated the mean point estimate, the  median absolute deviation (MAD)   and the  maximum likelihood estimates (MLE); the results are in Table \ref{t2} of Appendix B. Both algorithms are generally  more precise (with smaller MAD) than the MLE,  and  Algorithm 2 tends to be the most precise (with the smallest MAD) for small values of $\alpha$ and the sample size. The point estimates from both algorithms are very close  but  can be considerably different from the MLE even for relatively large data sizes.  It is observed that both  bias and variability  decrease  as the sample size increases in all cases as expected.

 
\section{Predictive distributions} 


Consider the exponential data model 
\begin{equation}
f( x | \theta ) \; =\;   c \cdot    e^{-a \cdot \theta \; + \;   b \cdot \log (\theta) }   , 
\end{equation}
where  $a=a(x), b =b(x),  c=c(x) \in \mathbb{R}^+.$   The prior predictive distribution of a conditionally independent new observation $x^*$ is 
\begin{equation}
p(x^*) \; = \; \int_{\theta} f( x^* | \theta ) \cdot IS_{\alpha, \rho} (\theta) d \theta  \; \sim \; c^*\;  \Gamma ( b^*+1)  \; \rho^{b^*} \; E_{\alpha,\alpha b^* + 1}^{b^*+1} (-a^*   \rho ),   \label{priorp}
\end{equation}

where  $a^*=a(x^*)$, $b^*= b(x^*)$,  and $c^*=c(x^*)$ is an appropriate  normalizing constant or function.  We now propose a class of prior predictive densities for a family of transformations $a(X^*)$ that provides a way of simulating $X^*$ by generating $A^*$ or inverting $a(X^*)$, and hence allows testing of prior beliefs.   Note that these predictive distributions can be easily plotted  using the maximum likelihood estimates of the hyperparameters and the Monte Carlo approximation   (\ref{mcif}).

\emph{Proposition 2.}    Let  $b$  and $c$  be    free of  $x$ (see Table \ref{t1}) and $X^* \to a(X^*)=A^*$  be a one-to-one mapping and invertible. Then  the prior predictive density of $A^*$  is  

\begin{equation}
p(a^*) \; = \;   b \;\rho \; \Gamma(1+\alpha (b -1) )  \;   E_{\alpha,\alpha b + 1}^{b+1} (-a^*   \rho).  \label{priorp2}
\end{equation}

\emph{Proof.} The proof follows from  the well-known Mellin transform of the generalized Mittag-Leffler function \citep{sap97}  below for $z=1, \eta=\alpha, \nu=\alpha b +1, \tau=b+1$ as 

\begin{equation}
\int_{\mathbb{R}^+} w^{z-1} \; E_{\eta,\nu}^\tau ( - \rho w) \; dw= \frac{\Gamma(z)\;\Gamma(\tau-z)}{\Gamma (\tau)  \; \rho^z \; \Gamma(\nu -z\eta)}.  \label{mellin1}
\end{equation} \qed

The above family of   proper density functions can easily be plotted  using the approximation in  (\ref{mcif}).  

\emph{Corollary 5.}   The $k$th moment can be calculated using formula (\ref{mellin1}) as 
\begin{equation}
\mathbf{E} \left(A^*\right)^k \;= \; \frac{  \Gamma(\alpha b +1 -\alpha) \; \Gamma( k+1) \; \Gamma(b+1-(k+1)) }{\Gamma(b) \;   \rho^{k} \; \Gamma(\alpha b +1 - (k+1)\alpha ) }.  \label{kmnt}   
\end{equation}
\emph{Proof.} It is trivial and is omitted. \qed

Formula (\ref{priorp2}) can be checked to be  proper  when $k=0$ in the preceding corollary.  The  $k$th moment exists provided $b>k$. 
  
\emph{Example 1.} Let $f( x | \theta ) = \theta \exp(-\theta x),  \in \mathbb{R}^+.$ Then  $a^*=x^*, b= 1,$ and thus,
\begin{equation}
 p(a^*)  \; = \;   \rho  \;E_{\alpha,\alpha + 1}^{2} (-x^*   \rho),    \quad \text{and} \quad   \mathbf{E} \left(A^*\right)^k \; = \; \mathbf{E} \left(X^*\right)^k \; = \;  \frac{ \Gamma( k+1) \; \Gamma (1-k) }{   \rho^{k} \Gamma (1 - \alpha k ) }, \; \; k \notin \mathbb{N}^+.
\end{equation}

 \emph{Example 2.} Let $f( x | \theta ) \sim \theta^{1/2}  \exp \left(-\theta x^2 \right), x \in \mathbb{R}$. Then  $a^*= x*^2 , b=1/2, $ and therefore,
\begin{equation}
p(a^*)  \; = \;   \rho^{1/2} \; (1/2)   \;E_{\alpha,\alpha/2 + 1}^{3/2} (-a^*   \rho); \; \;  \mathbf{E} \left(A^*\right)^{k} = \mathbf{E} \left(X^*\right)^{2k} =  \frac{\Gamma(1-k)\; \Gamma (k+1/2)}{\sqrt{\pi} \; \Gamma ( 1-\alpha k)  \; \rho^k}, \; k \notin \mathbb{N}^+. 
\end{equation} 

When $k=0$,  the preceding  $p(a^*) $'s   are easily checked as proper prior predictive density functions.   

After observing the data $\bm{x} \in \mathbb{R}^n$,  the posterior predictive distribution of a conditionally independent new observation $A^*=a(X^*)$ is 
 \begin{equation}
 p(a^* | \bm{x}) \; = \; \int_{\theta} f( a^* | \theta, \bm{x} ) p (\theta | \bm{x}) d \theta  \; \sim \;        c^*\; \rho^{b^*}      E_{\alpha,\alpha b' + 1}^{b'+1} \left(- (a+a^*)  \rho \right) , \qquad   b'=b^* + b,  \label{predpost}
\end{equation}
 where    $ a=a(\bm{x}), b' \in \mathbb{R}^+,  b\text{ and $b^*$-both known and free of} \; x\; (\text{like most of the examples in Table \ref{t1}}),$  and $c^*=c(x^*) \in \mathbb{R}^+$ is some normalizing function or constant.

\emph{Example 3.} Let $f( x | \theta ) = \theta \exp(-\theta x), x\in \mathbb{R}^+.$ Then  $a^*=x^*, b^*=1, a=\sum x_i,  b=n, $ and thus,
\begin{equation}
 p(a^* | \bm{x})  \; \sim \; \rho \;       E_{\alpha,\alpha (n+1) + 1}^{n+2} (- (a+a^*)    \rho).
\end{equation}

 \emph{Example 4.} Let $f( x | \theta ) \sim \theta^{1/2}  \exp \left(-\theta x^2 \right), x \in \mathbb{R}$. Then  $a^*= x*^2 , b^*=1/2,  a=\sum x_i^2,  b=n/2$ and hence,
\begin{equation}
p(a^* | \bm{x})  \; \sim \;    \rho^{1/2}  \;   E_{\alpha,\alpha (n+1)/2 + 1}^{(n+1)/2+1} (- (a+a^*)     \rho).
\end{equation} 
 
The above  posterior predictive distributions  can directly be plotted   using (\ref{mcif}) also. 


\section{Hierarchical models}
 
Recall that $p( \lambda | data) \sim p( data | \lambda) \cdot p( \lambda)$, where   the marginal prior $p( \lambda) \sim \int  p( \lambda | \theta) IS_{\alpha,\rho} (\theta) d\theta \sim  \int  \exp (- g(\lambda)  \cdot \theta) \theta^b IS_{\alpha,\rho} (\theta) d\theta    \sim    E_{\alpha,\alpha b + 1}^{b+1} (- g(\lambda) \cdot \rho )$ follows from \emph{Corollary 1},   with the inverse stable density as the mixing distribution.  The marginal prior family can  be   plotted using the Monte Carlo algorithm above and can be made proper  using the Mellin transform property (\ref{mellin1}). Below are some applications of the inverse stable prior  in hierarchical   contexts.

\noindent \textbf{Normal-normal model}

Let $\theta$ be the inverse variance and consider 

 \begin{equation}
\overline{X} \; | \;  \lambda \stackrel{d}{=}  N (\lambda, 1/n), \quad \Lambda \; | \; \theta \stackrel{d}{=}  N\left( 0, \; \theta^{-1} \right), \quad \theta \stackrel{d}{=}  IS_{\alpha, \rho} \left(  \theta \right), \; \; \;\;  \alpha, \rho\text{-known.}
\end{equation}
The marginal prior of $\Lambda$  can be shown as 
\begin{equation}
p(\lambda) \;\sim  \;    \; \rho^{1/2}\;  E_{\alpha,\alpha/2 + 1}^{1/2+1} (- \lambda^2 \rho/2 ).  
\end{equation}
Also, the hyperparameters $\alpha$ and $ \rho$ can be estimated by  maximizing
 \begin{equation}
data \;  |  \; \alpha, \rho \; \sim  \;   \mathbf{E}_\Theta \left[  \left(2 \pi \frac{n+\Theta}{n\Theta} \right)^{-1/2}  \exp \left( -\frac{ n \Theta \overline{x}^2}{2(n+\Theta) } \right) \right].
\end{equation}
The Gibbs sampler for this model can be performed  as
\begin{equation}
\Lambda \;  |  \;data, \theta \; \stackrel{d}{=}  \; N\left( \frac{\overline{x}}{1+\theta/n }\; ,\; \left(n+\theta \right)^{-1} \right),
\end{equation}
and
\begin{equation}
\Theta \; | \; \lambda, data \; \sim\; \theta ^{1/2} \exp \left( -\frac{\lambda^2}{2} \theta \right) \cdot  IS_{\alpha, \rho} \left(  \theta \right).
\end{equation}

\noindent \textbf{Poisson-exponential model}

Let 

\begin{equation}
data \; | \;  \lambda \stackrel{d}{=}  {\tt Poisson} (\lambda), \quad \Lambda \; | \; \theta\stackrel{d}{=}  \theta e^{-\lambda \theta} ,\quad \Theta  \stackrel{d}{=}  IS_{\alpha, \rho} \left(  \theta \right), \; \; \;  \alpha, \rho\text{-known.}
\end{equation}

The marginal prior of $ \Lambda$  can be obtained as 
\begin{equation}
p(\lambda) \;  \sim \;    \; \rho \;  E_{\alpha,\alpha + 1}^{2} \left(- \rho \lambda  \right).
\end{equation}

Moreover,  maximizing
 \begin{equation}
data \;  |  \; \alpha, \rho \; \sim  \;   \mathbf{E}_\Theta \left[  \frac{\Gamma \left( \sum_{j=1}^n x_j +1 \right) }{ \left( n+ \Theta\right)^{\sum_{j=1}^n x_j  +1} \; \prod_{j=1}^n x_j }\right] \label{pgcon}
\end{equation}
gives estimates of the hyperparameters $\alpha$ and $ \rho.$ The Gibbs sampler for this model can be accomplished as
\begin{equation}
\Lambda \;  |  \;data, \theta \; \stackrel{d}{=}  \;   {\tt Gamma} \left( \sum_{j=1}^n x_j + 1 \; ,\;  n+\theta  \right),
\end{equation}
and
\begin{equation}
\Theta \; | \; \lambda, data \; \sim\; \theta \exp \left( -\lambda \theta \right) \cdot  IS_{\alpha, \rho} \left(  \theta \right).
\end{equation}

\noindent \textbf{Poisson-exponential with unequal intensity rates}

Let 
\begin{equation}
X_i \; | \;  \lambda_i \stackrel{d}{=}  {\tt Poisson} (\lambda_i), \quad \Lambda_i \; | \; \theta \stackrel{d}{=}   \theta e^{-\lambda_i \theta}, \quad \Theta \stackrel{d}{=}  IS_{\alpha} \left(  \theta \right) \;=\; IS_{\alpha, 1} \left(  \theta \right),\;\; \; \alpha\text{-known}.
\end{equation}

The marginal prior of $\bm{\Lambda}= \left(\Lambda_1, \Lambda_2,\ldots, \Lambda_n\right)'$  can be calculated as 
\begin{equation}
p(\bm{\lambda}) \;  \sim  \;    E_{\alpha,\alpha n + 1}^{n+1} \left(-  \sum_{i=1}^n \lambda_i  \right), \qquad \bm{\lambda} = (\lambda_1,\lambda_2, \ldots, \lambda_n)'.
\end{equation}
Then
\begin{equation}
\Lambda_i \;  |  \;  data,    \theta \; \stackrel{d}{=} \;  {\tt Gamma} \left( x_i + 1 \; ,\;  1+\theta  \right),
\end{equation}
and
\begin{equation}
\mathbf{E} \left( \Lambda_i \;  |  \;  data,    \theta  \right)  \; =  \;( x_i + 1) /(1+\theta)=(1-\kappa)x_i + \frac{1}{\theta} \kappa,
\end{equation}
where $\kappa=\theta/(1+\theta)$ describes the shrinkage.  Hence,
\begin{equation}
\mathbf{E}\left( \Lambda_i \;  |  \;  data  \right)  \; = \;( x_i + 1)  \mathbf{E}_{\Theta |data}[  (1+\theta)^{-1} ], 
\end{equation}
where
\begin{equation}
\Theta \; |\; data \;  \stackrel{d}{=}  K  \; \left( \frac{1}{1+\theta}\right)^{\sum_{j=1}^nx_j} \left(1- \frac{1}{1+\theta}\right)^n   IS_\alpha(\theta) = K  \; \left(1-\kappa\right)^{\sum_{j=1}^nx_j} \kappa^n \; IS_\alpha(\theta). 
\end{equation}
The normalizing constant $K$ can be straightforwardly calculated using Monte Carlo integration.    Therefore,
\begin{eqnarray}
\mathbf{E}\left( \Lambda_i \;  |  \;  data  \right) = ( x_i + 1)  \cdot \mathbf{E}_\Theta\left[   K  \; \left(1-\kappa\right)^{\sum_{j=1}^nx_j+1} \kappa^n  \right].
\end{eqnarray}


\noindent \textbf{Global shrinkage in normal-normal model}

Let $\theta$ be the  inverse variance.  
Consider

\begin{equation}
X_i \; | \;  \lambda_i \stackrel{d}{=}  N(\lambda_i, 1), \quad \Lambda_i \; | \; \theta \stackrel{d}{=}  N (0,  \theta^{-1}), \quad \Theta \stackrel{d}{=}  IS_{\alpha} \left(  \theta \right), \; \;   \alpha\text{-known}. \label{is1}
\end{equation}

The marginal prior for $\bm{\Lambda}$ is 
\begin{equation}
p(\bm{\lambda}) \;  \sim \;      E_{\alpha,\alpha n/2 + 1}^{n/2+1} \left(-   || \bm{\lambda}||_2^2/2  \right) .
\end{equation}

Then
\begin{equation}
\Lambda_i \;  |  \;  data,    \theta \; \stackrel{d}{=} \;  N \left( \frac{x_i}{1+\theta} \; ,\;  \frac{1}{1+\theta}  \right),
\end{equation}
and
\begin{equation}
\mathbf{E} \left( \Lambda_i \;  |  \;  data,    \theta  \right)  \; =  \; x_i /(1+\theta)=(1-\kappa)x_i,
\end{equation}
where $\kappa= \theta /(1+\theta).$   Hence,
\begin{equation}
\mathbf{E}\left( \Lambda_i \;  |  \;  data  \right)  \; = \; x_i  \cdot \mathbf{E}_{\Theta |data}[  (1+\theta)^{-1} ], 
\end{equation}
where
\begin{eqnarray}
\Theta\; | \; data \; &  \stackrel{d}{=}  & K  \; \left( \frac{1}{1+\theta}\right)^{n/2} \exp \left(    || \bm{x}||_2^2/[2(1+\theta)] \right)   \;\theta^{n/2}\; IS_\alpha(\theta) \\
&= &  K  \; \kappa^{n/2} \exp \left( (1-\kappa)  || \bm{x}||_2^2/2\right)  \; IS_\alpha(\theta).
\end{eqnarray}
 Thus,
\begin{eqnarray}
\mathbf{E}\left( \Lambda_i \;  |  \;  data  \right) =  x_i \cdot \left(1 -   \mathbf{E}_\Theta\left[ K  \; \kappa^{n/2 +1} \exp \left( (1-\kappa) || \bm{x}||_2^2/2 \right)   \;\right] \right).
\end{eqnarray}

\begin{figure}[h!t!b!p!]
     \centering
                         \includegraphics[height=3in, width=6in]{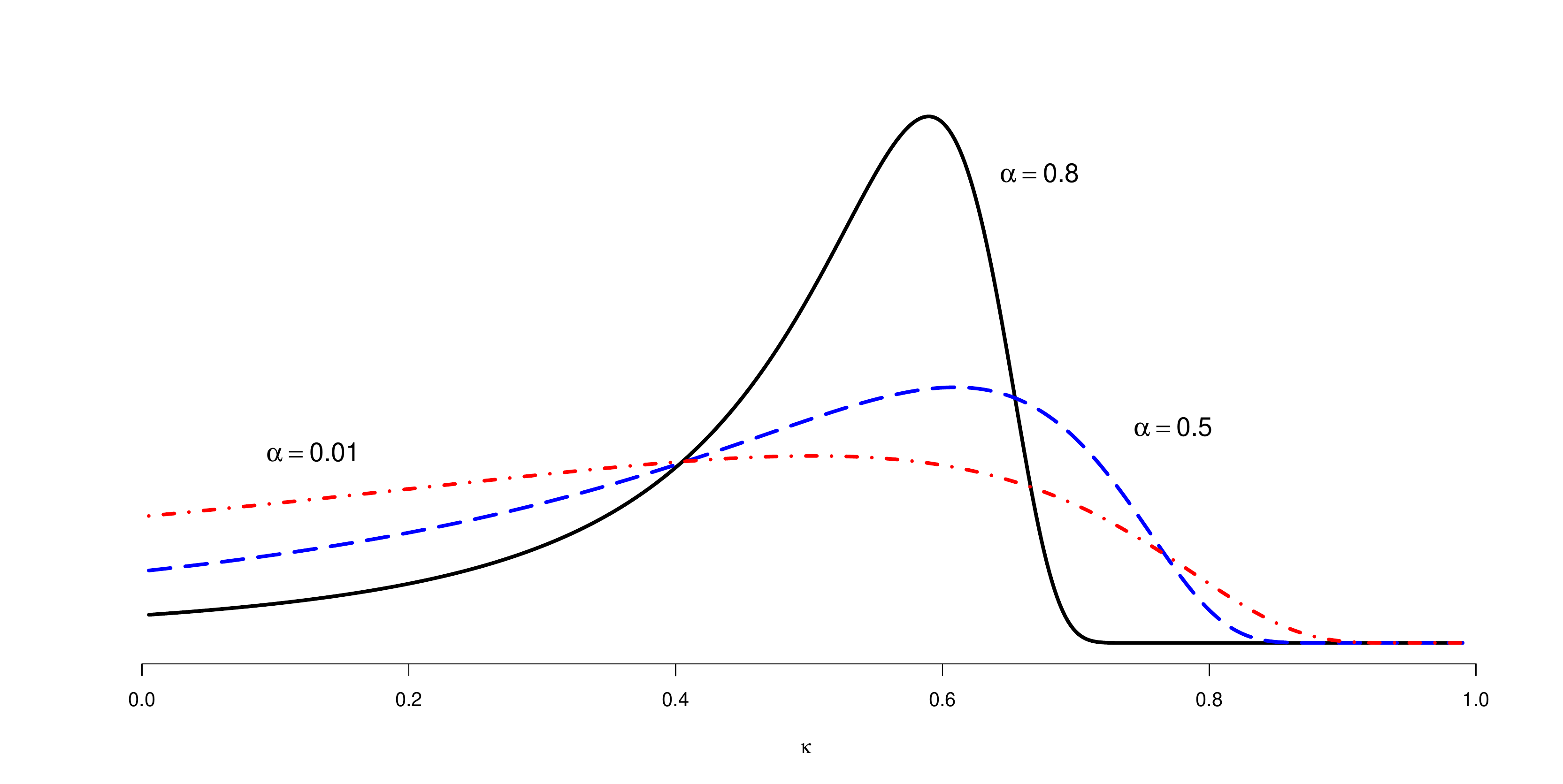}
       \renewcommand\abovecaptionskip{0pt}
     \renewcommand\belowcaptionskip{0pt}
 \caption{Marginal prior  densities of  $\kappa$ as a function of  $\alpha$.}
\label{f12}
\end{figure}

The implied density for $\kappa$  is $ p(\kappa) \sim \left( 1-\kappa\right)^{-2}\cdot IS_\alpha \left( \frac{\kappa}{1-\kappa} \right)$ \citep[see also][]{pas12a}.  The effect of $\alpha$  on the density of  $\kappa$ is  shown in the figure below.  The density converges to dirac measure at $\kappa= 1/2$ as $\alpha \to 1^-$ and  $ p(\kappa) \sim  1/\Gamma (1-\alpha )$  as $\kappa \to 0^+.$   From Figure \ref{f12}, the density  $p (\kappa)$ is bounded  at  both endpoints  $(\kappa=0 \; \text{or}\; \theta =0 \;   ; \; \kappa=1 \; \text{or}\; \theta = \infty )$     

If we consider the normal scale mixture in application 4 of section 1, then it can be shown that 
\begin{eqnarray}
\mathbf{E}\left( \Lambda_i \;  |  \;  data  \right) =  x_i \cdot \left[1 -   \mathbf{E}_{\delta^2 | data} \;  \left( \kappa' | \; data \;\right) \right]
\end{eqnarray}
where $\kappa'=1/(1 + \delta^2).$   

We compare the two priors   in terms of their observed frequentist risks ($L_1$- and squared $L_2$- norms).   We generated 1000 data samples following these five   setups: 

\emph{Case I:} $\Lambda_i \sim  N(0, 1)$ and  $X_i \; | \;  \lambda_i \sim N(\lambda_i, 1), \; i=1,\ldots, 9.$


\emph{Case II:}   $   \Lambda_1, \Lambda_2, \Lambda_3  \sim N(0, 1); $ $X_{1:3} \; | \;  \lambda_1 \sim  N(\lambda_1, 1), X_{4:6} \; | \;  \lambda_2 \sim N(\lambda_2, 1), X_{7:9} \; | \;  \lambda_3 \sim N(\lambda_3, 1).$

\emph{Case III:}     $   \Lambda_1, \Lambda_2, \Lambda_3  \sim N(0, 100); $ $X_{1:3} \; | \;  \lambda_1 \sim  N(\lambda_1, 1), X_{4:6} \; | \;  \lambda_2 \sim N(\lambda_2, 1), X_{7:9} \; | \;  \lambda_3 \sim N(\lambda_3, 1).$

 \emph{Case IV:}  $ X_i \sim N(\lambda_i,1),  \;  \Lambda_i  \sim   0.1\times    t(df=2, \; scale=3) + 0.9\times \delta_0$ where $\delta_0$ is the Dirac measure at zero; $i=1,\ldots, 250.$ 

\emph{Case V:} $ X_i \sim N(\lambda_i,1.5),  \;  \Lambda_i \sim   0.1\times    t(df=2, \; scale=3) + 0.9\times \delta_0;  i=1,\ldots, 250.$ 
 
Table \ref{t3} summarizes the results over the 1000 samples.  For each of the five cases and procedures (inverted beta, inverse stable with $\alpha=0.01, 0.5, 0.99$) we  present 
\begin{equation}
\frac{ \sum_{j=1}^{1000}\sum_{i=1}^n |\lambda_{ji}- \hat{\lambda}_{ji} |^k}{1000}, \qquad k=1,2,
\end{equation}
where  $\lambda_{ji}$s are the selected values used for comparison with the posterior means,  and $n$ is the number of $\lambda_i$s for that case.  Overall, the results are similar  for both priors, suggesting the value of further investigation of the inverse stable family of priors as an alternative to the inverse beta. In terms of the $L_1$ risk, we are able to find  $\alpha$'s  where the inverse stable provides a prior at least as good as the inverse beta.  For the squared $L_2$ risk, the two priors give similar results for   Cases I, II, and V.   Note that  the posteriors were standardized   using numerical (inverted beta) and Monte Carlo methods.     We subjected $2\times 10^7$ random numbers for acceptance or rejection in our simulations.

\begin{table}[h!t!b!p!]
\caption{\emph{Mean $L_1$ and squared $L_2$ risks. }} \centerline {
\begin{tabular*}{7.1in}{c|cccccccc}
  &  \multicolumn{4}{c}{$L_1$} & \multicolumn{4}{c}{Squared $L_2$}  \\
 \hline
  &  Inv Beta  & \multicolumn{3}{c}{Inv Stable} &   Inv Beta  & \multicolumn{3}{c}{Inv Stable} \\
& & $\alpha=0.01$ & $\alpha=0.5$ & $\alpha=0.99$   & & $\alpha=0.01$ & $\alpha=0.5$ & $\alpha=0.99$  \\
\hline
Case I &6.247 &  \bf{6.010} & 6.046 & 6.147& 6.721& \bf{6.226}& 6.284 & 6.425\\
Case II & \bf{7.386} &  7.642& 7.567& 7.546& 10.655 & 11.217& 10.952& \bf{10.520} \\
Case III & 66.787 & \bf{65.768} & 66.118& 66.085& 1068.171 &  \bf{981.817} &  993.494&  993.166 \\
Case IV & 158.674 & 156.490 & \bf{156.378}& 155.814 &  \bf{181.981}& 201.705 & 201.712 & 202.071\\
Case V & 245.558 & 235.482 & 235.386 & \bf{234.689} &  \bf{386.416}  &389.802 & 389.568 & 387.966\\
\hline 
\end{tabular*}
}
  \label{t3}

\end{table}
 
\begin{figure}[h!t!b!p!]
     \centering
\includegraphics[height=3in, width=3.4in]{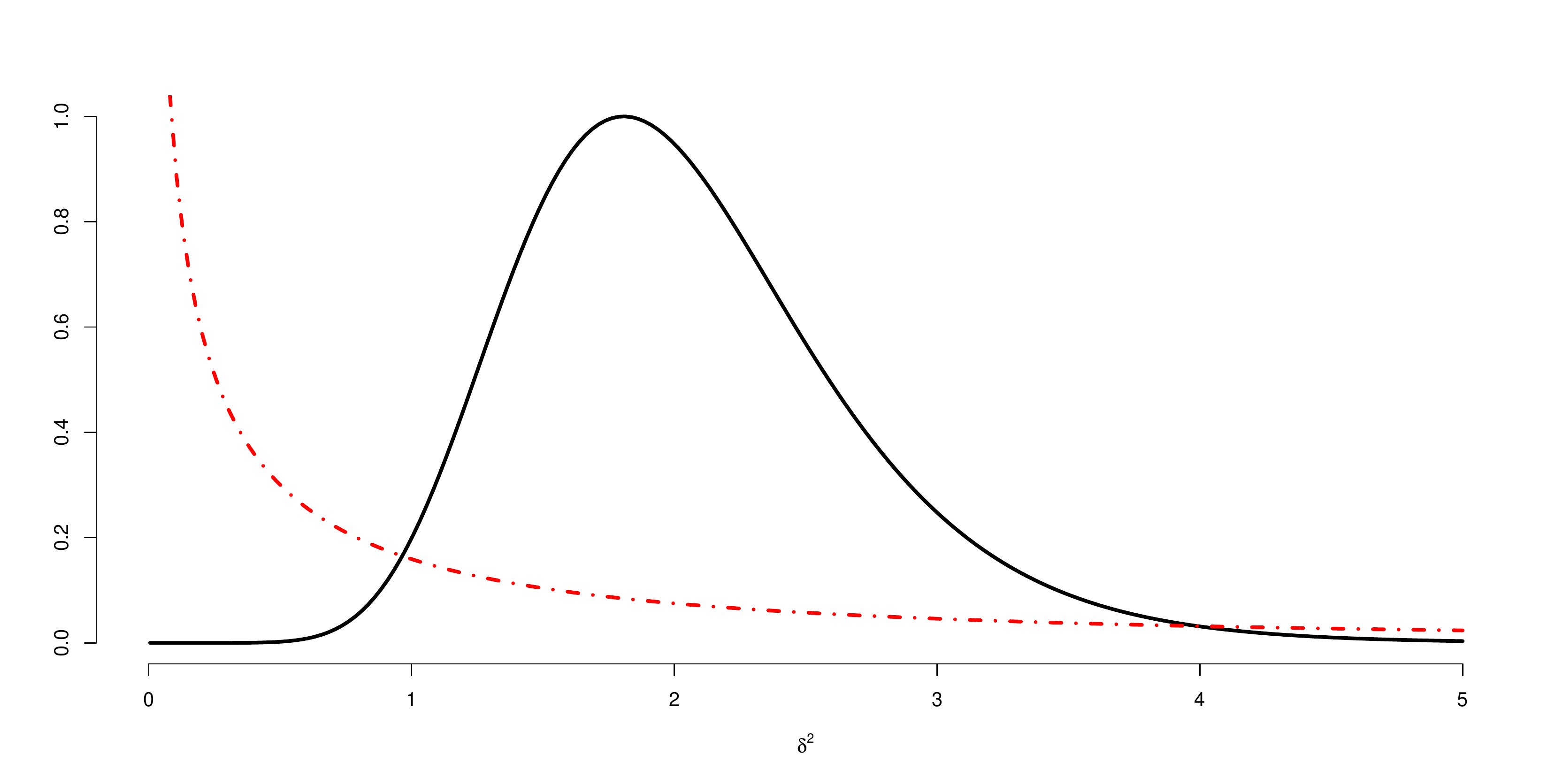}
\includegraphics[height=3in, width=3.4in]{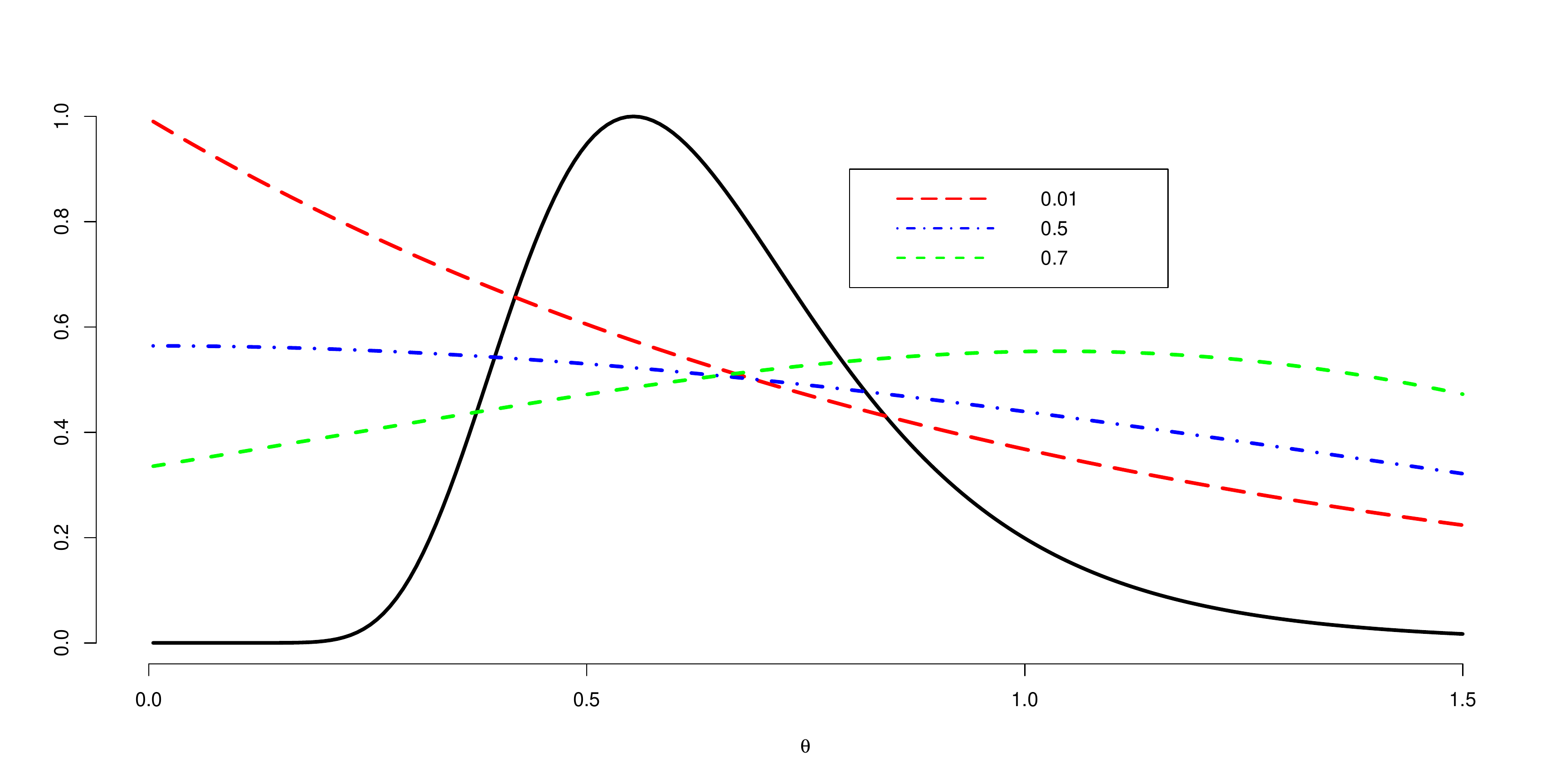}


 \caption{
 The solid black lines show the  likelihoods as a function of $\delta^2$ and $\theta$, respectively.   (Left) The solid black line  shows the  likelihood  as a function of $\delta^2$  together with the  inverted beta prior. (Right)   The black line represents the  likelihood as a function of  $\theta$ together with the inverse stable priors  for $\alpha=0.01(red),  0.5 (blue), 0.7(green).$}  \label{f6}
\end{figure}

 For several data sets, we have plotted in Figure \ref{f6} the likelihood of $\delta^2$ based on (\ref{ib}) against the inverted beta prior in (\ref{ib2}) .

\pagebreak

As expected, the inverted beta priors are usually informative. That is, the prior  density has appreciable mass where the likelihood is negligible, typically for small values of $\delta^2.$ The corresponding plots of the  likelihood of  $\theta$  against the inverse stable prior in (\ref{is1}) have much better behavior.  Fifty data points were generated  from $N\left(0, \;1.5^2\right)$ and the corresponding likelihoods were normalized to  obtain a maximum of one.


\noindent \textbf{Local shrinkage in normal-normal model}

Let  $\theta_i$ be the local inverse variance.  When  $\Lambda_i \; | \; \theta_i \stackrel{d}{=}  N \left(0, \theta_i^{-1} \right)$ and  $\Theta_i \;  \stackrel{d}{=}  IS_\alpha (\theta_i)$  above, then the marginal prior for $\Lambda_i$ is 
\begin{equation}
p(\lambda_i) \;  \sim \;      E_{\alpha,\alpha/2 + 1}^{1/2+1} \left(-   \lambda_i^2/2  \right) .
\end{equation}

Furthermore, 
\begin{equation}
\Lambda_i \;  |  \;  x_i,    \theta_i \; \stackrel{d}{=} \;  N \left( \frac{x_i}{1+\theta_i} \; ,\;  \frac{1}{1+\theta_i}  \right),
\end{equation}
and
\begin{equation}
\mathbf{E} \left( \Lambda_i \;  |  \;  x_i,    \theta_i  \right)  \; =  \; x_i /(1+\theta_i)=(1-\kappa_i)x_i, \qquad \kappa_i = \theta_i /(1+\theta_i).
\end{equation}
Therefore,
\begin{equation}
\mathbf{E}\left( \Lambda_i \;  |  \;  x_i \right)  \; = \; x_i  \cdot \mathbf{E}_{\Theta_i |x_i}[  (1+\theta_i)^{-1} ], 
\end{equation}
where
\begin{eqnarray}
\Theta_i \; | \; x_i \; &  \stackrel{d}{=}  & K  \; \left( \frac{1}{1+\theta_i}\right)^{1/2} \exp \left(  x_i^2/[2(1+\theta_i)] \right)  \; IS_\alpha(\theta_i) \; \theta_i^{1/2}  \\
&= &  K  \; \kappa_i^{1/2} \exp \left( (1-\kappa_i ) x_i^2/2\right)  \; IS_\alpha(\theta_i)  .
\end{eqnarray}
  Hence,
\begin{eqnarray}
\mathbf{E}\left( \Lambda_i \;  |  \; x_i  \right) =  x_i \cdot \left( 1 - \mathbf{E}_{\Theta_i}\left[ K  \; \kappa_i^{1/2 +1} \exp \left( (1-\kappa_i) x_i^2/2 \right)   \right] \right).
\end{eqnarray}

 \section{Real data illustration} 
 
We  considered the Quine data (available in the \texttt{MCMCpack} package of R). The Quine dataset has $n=146$ children from Walgett, New South Wales, Australia, who were classified by culture, age, sex and learner status,  and the number of days absent ($X_j$'s) from school in a particular school year was recorded.  We  applied the hierarchical  model below only to  the number of days absent:
\begin{eqnarray}
X_j \; |  \; \lambda &\sim& \; {\tt Poisson} (\lambda),  \\
\Lambda \; | \;  \theta  &\sim& \;  \theta e^{-\lambda \theta}, \\
\Theta \; | \; \alpha, \rho \;   &\sim& \; IS_{\alpha,  \rho} (\theta) , \\
(\alpha, \rho)  \;   &\sim& \;  h(\alpha) \; g_{\alpha} (\rho ),
\end{eqnarray}
where    $h(\cdot)$  is a well-defined function in  $(0,1)$ and  $g_{\alpha} (\rho)$ is the heavy-tailed (with infinite mean) $\sqrt{\alpha}^{\; +}$-stable hyperprior in $\mathbb{R}^+.$ Note that we allow dependence between hyperparameters.    Moreover,    $\sum_{j=1}^{n} X_j = W\stackrel{d}{=}{\tt Poisson} (n \lambda)$ and 

\begin{eqnarray}
W \; \big{|} \;  \alpha, \rho  \; &\sim& \; \int \left[ \int \frac{e^{-n \lambda} (n\lambda)^w}{w!} \cdot \theta e^{-\lambda \theta} \cdot IS_{\sqrt{\alpha},  \rho^{-\alpha}}(\theta) \; d \lambda \right]  \; d\theta \\
\; &=& \; \int \left[ \int  \frac{n^w \; e^{-(n + \theta) \lambda}  \lambda^w}{w!}    \;  d \lambda \right]  \theta \cdot  IS_{\sqrt{\alpha},  \rho^{-\alpha}}(\theta) \; d\theta \\
\; &=& \;  \mathbf{E}_\Theta  \left[  \frac{   \Theta \; n^{ w}}{ (n+\Theta)^{w+1} }  \right].  \label{pgcon2}
\end{eqnarray}
A simple grid search  (or  any other optimization method) can straightforwardly be applied to (\ref{pgcon2})   to find the maximum likelihood estimates of the hyperparameters  using  (\ref{mcif}).  Observe   that the parameter $\alpha$ is bounded in $(0,1)$, making the maximum likelihood search relatively fast.        
 
Let $\eta =(\alpha, \rho)$. Then a Gibbs sampler consists of the following:

\begin{eqnarray*}
\Lambda \; | \; \theta, w, \eta \;  &\sim& \;  \Gamma (w+1, n+\theta), \\
\Theta \; | \; \lambda, w, \eta  \;  &\sim& \; \theta e^{-\lambda \theta}  IS_{\alpha,  \rho}(\theta),\\
\eta \; | \; \lambda, w, \theta  \;  &\sim& \;  IS_{\alpha,  \rho}(\theta) h(\alpha)   g_{\alpha} (\rho ).
\end{eqnarray*}
 
Using a non-informative prior $h(\alpha)=1,$  we generated 13000 joint posterior samples with 3000 observations as burn-ins. Note that $g_{\alpha} (\rho )$ is heavy-tailed  where the mean is non-existent \citep[see][and the references therein for robustness]{ bdpw16}. Furthermore, samples from the first two conditional distributions of the Gibbs sampler above can directly be obtained using a built-in function (in R, for example) and  Algorithm 2, correspondingly.  We utilized a grid-based method to sample from   $(\eta \; | \; \lambda, w, \theta)$  using  at least 30000  equally spaced (marginally)  $(\alpha, \rho)$ values   from $(0.01 ,0.99) \times (0.005 ,3)$  as grids.  In this process, we evaluated  $ IS_{\alpha,  \rho}(\theta) h(\alpha)   g_{\alpha} (\rho )$ using the following widely used  integral formula:
\begin{equation}
g_{\alpha}(s) = \frac{\alpha s^{1/(\alpha-1)}}{\pi (1-\alpha)}
\int\limits_{-\pi/2}^{\pi/2} \exp \left\{ -s^{\alpha/(\alpha-1)}
U(\phi; \alpha)\right\} U(\phi; \alpha) d\phi, 
\end{equation}
where
\begin{equation}
U(\phi; \alpha) = \left [ \frac{\sin (\alpha(\phi+\pi/2))}{\cos
\phi} \right]^{\alpha/({\alpha-1})} \frac{\cos \left( (\alpha-1)\phi
+ \alpha \pi/2 \right)}{\cos \phi},
\end{equation}
and normalized the values by their sum to form the weights used in sampling.

\begin{figure}[h!t!b!p!]
     \centering
\includegraphics[height=3in, width=3.5in]{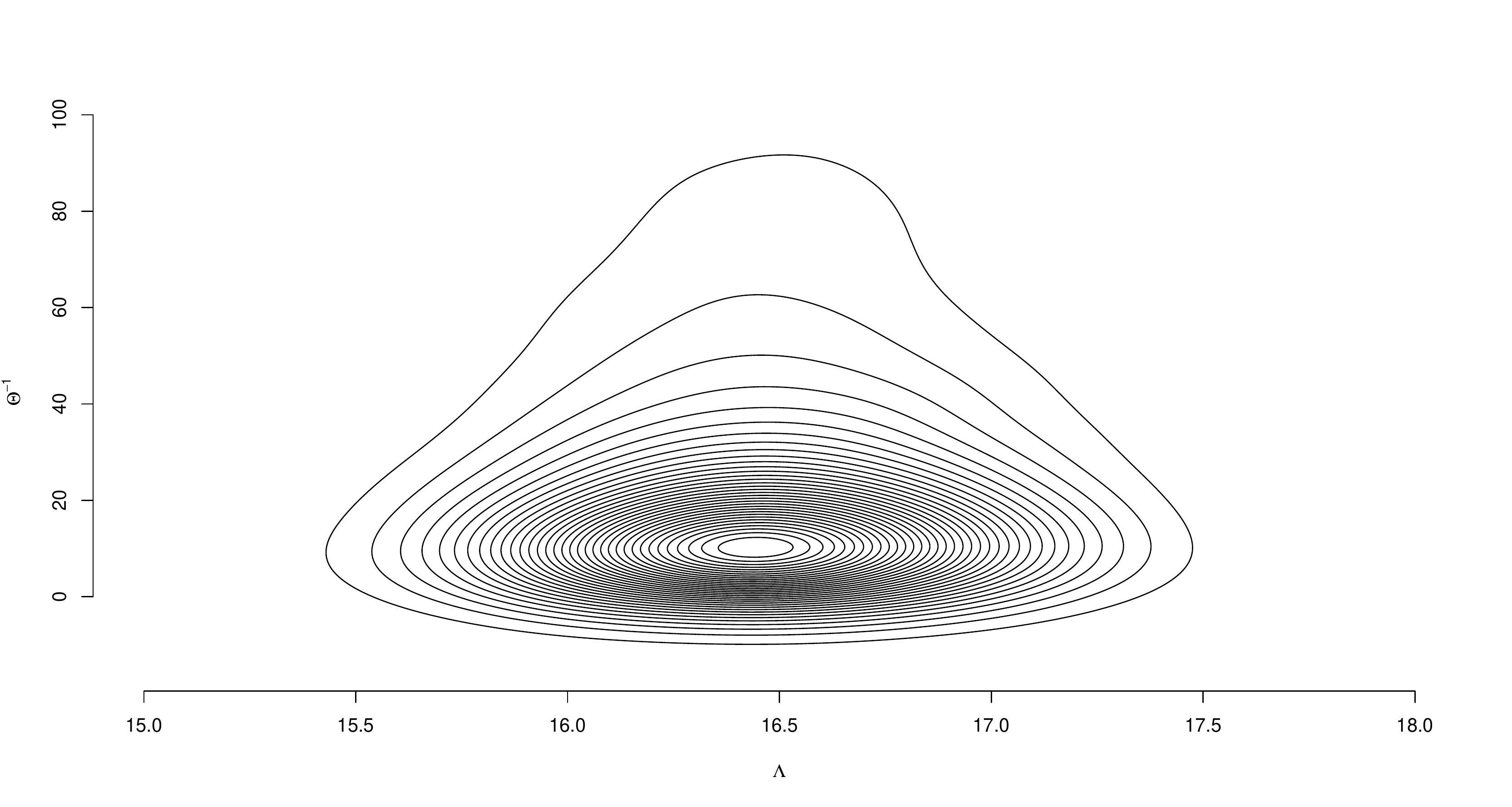}
\\
 \caption{Joint posterior fit of  $(\Lambda ,\Theta^{-1}).$  }  \label{f3}
\end{figure}

 The simulated posterior fit of  $(\Lambda, \Theta^{-1})$  (means)    is in Figure \ref{f3}.

 The point estimates and  highest posterior density (HPD)  credible intervals  for $\lambda$ and $\theta^{-1}$     are in Table \ref{t5}.  Note that  the \texttt{HPDinterval} function from the \texttt{lme4}  package  of R  was utilized for this exercise.     The  point estimates and the  95\% HPD credible intervals for the {\tt Poisson} rate $\lambda$ tend to cluster around $16.456$ and $(15.796\;, 17.108)$, respectively.  The posterior distribution of $\theta^{-1}$  is   right-skewed  where estimates range from 1.921 to 116.488.  Notice that the estimates  for $\theta$ are easily obtained as  the reciprocals of the point estimate and credible interval for  $\theta^{-1}$.    

\begin{table}[h!t!b!p!]
\caption{\emph{Point estimates and 95\% HPD credible  intervals for the Quine data.  }} \centerline {
\begin{tabular*}{3.9in}{ c|c||c|c}
  \multicolumn{2}{c||}{$\lambda$}  &  \multicolumn{2}{c}{$\theta^{-1}$}    \\ 
\hline 
    Point   &    Interval  &    Point   &    Interval  \\
\hline \hline
       16.456& (15.796 , 17.108) &   32.748 & (1.921 ,  116.488)  \\   
\end{tabular*}
}
  \label{t5}
 \end{table}

\begin{figure}[h!t!b!p!]
     \centering
\includegraphics[height=3in, width=3.5in]{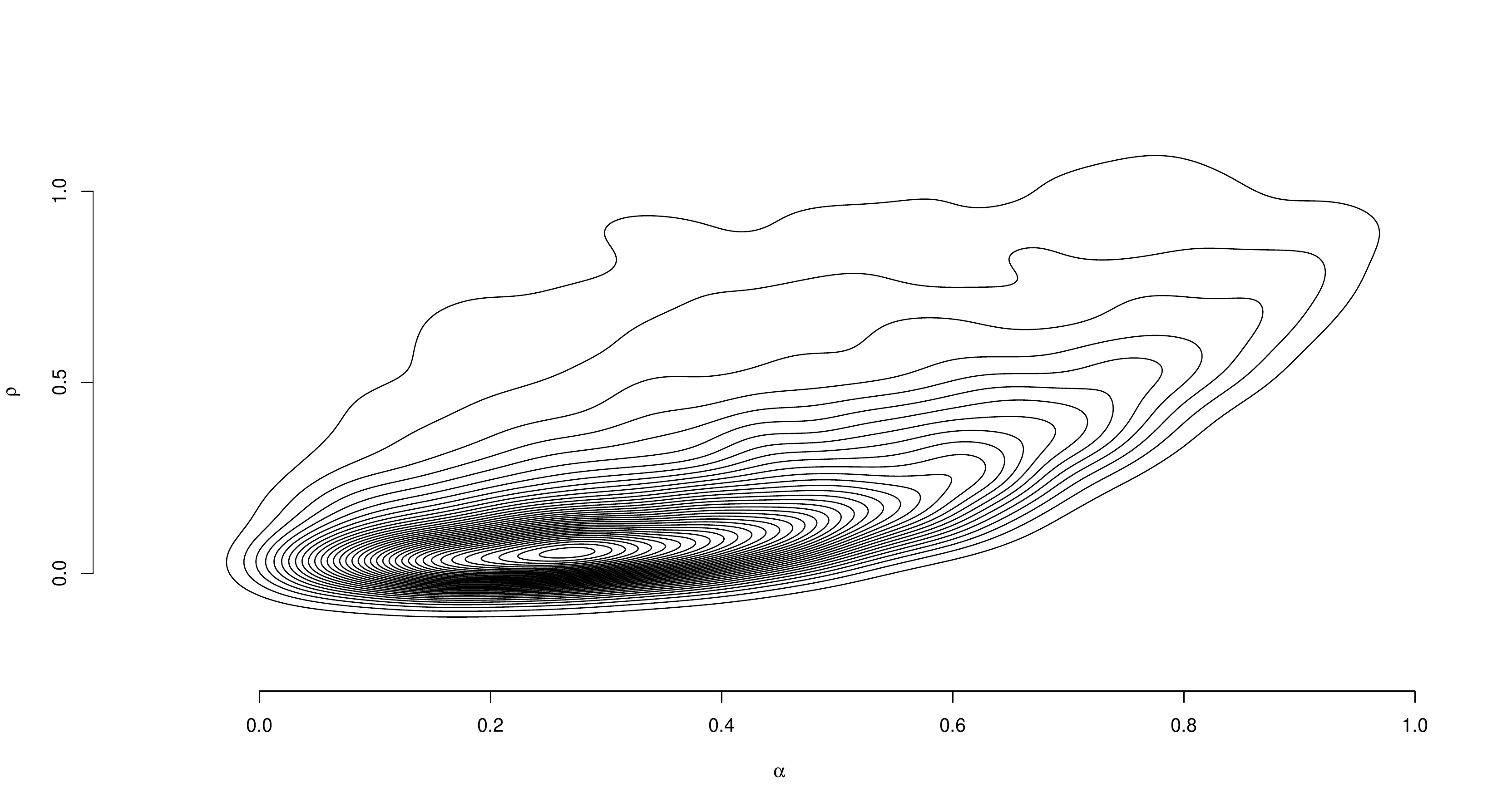}
\\
 \caption{Posterior  of  $(\alpha ,\rho).$  }  \label{f4}
\end{figure}

Moreover,  the point estimates and  credible intervals  for $\alpha$ and $\rho$   are in Table \ref{t5b}.   The  point estimates and the  95\% HPD  credible intervals for   $\alpha$ tend to cluster around $0.409$ and $(0.049\;, 0.804)$, respectively.  The  estimates  of $\rho$  range from 0.005  to  1.184.       These results are made apparent by the posterior plot of  $(\eta \; | \; \lambda, w, \theta)$ in Figure \ref{f4} .

\begin{table}[h!t!b!p!]
\caption{\emph{Point estimates and 95\% HPD credible  intervals for $\alpha$ and $\rho$.  }} \centerline {
\begin{tabular*}{3.5in}{ c|c||c|c}
  \multicolumn{2}{c||}{$\alpha$}  &  \multicolumn{2}{c}{$\rho$}    \\ 
\hline 
    Point   &    Interval  &    Point   &    Interval  \\
\hline \hline
       0.409& (0.049 , 0.804) &   0.351 & (0.005 ,  1.184)  \\   
\end{tabular*}
}
  \label{t5b}
 \end{table}

\section{Concluding points}
 
We propose  a family of proper and non-conjugate priors for an array of discrete and continuous exponential probability models. The resulting posterior family is  proper. The normalized posterior family then allows for the derivation of some properties such as the moment generating function, which leads to the closed-form formulas  of the moments including  the Bayes estimator with respect to the quadratic loss  criterion. The posterior class  ranges from a non-degenerate  to a degenerate  family of distributions. The parameter $\alpha \in (0 , \;1)$ modulates the   shapes of the posterior distribution   that adapts to a large class of signal patterns. We develop algorithms for calculating the Bayes point and interval estimates (under the squared error loss)  and for generating samples from the  posterior distributions. We  illustrate  the proposed methodology using a real data set and extend it to the hierarchical settings. In particular,  we show that inverse stable can provide better shrinkage than the inverted beta prior for the normal-normal model with a global variance.   We also  introduce a hyperprior density  of the form $h(\alpha) g_{\alpha} (\rho).$  

 From (\ref{mcif}) and \emph{Corollary 1},   maximum likelihood search can be made faster by  the boundedness of the parameter $\alpha$.   If  $p( data | \alpha, \rho)  \sim  \int  \int   p( data | \lambda)  p( \lambda | \theta) IS_{\alpha,\rho} (\theta) d\theta  d \lambda$ is well-behaved then point estimates of the  hyperparameters are easily obtainable.   

Note  that it is straightforward to specify a heavy-tailed  prior for $\Theta$.  Without loss of generality, let $\rho =1$ and 
\begin{equation}
\Theta \; \sim  \; \theta^{-1 -\alpha'/\alpha}\cdot  IS_{\alpha} \left(  \theta \right),  \label{htp}
\end{equation}
where $ 0 <\alpha \leq \alpha' < 1.$  By simple substitution,  it is easy to show that the mean of  $\Theta$ fails to exist in this case.  From Proposition 1,
\begin{equation}
p(\theta |  \bm{x}  ) \;  =  \;  \frac{e^{-a \cdot \theta \; + \; b' \cdot \log (\theta)}  \;  IS_{\alpha} \left(  \theta   \right) }{\Gamma ( b'+1)  \; \rho^{b'} \; E_{\alpha,\alpha b' + 1}^{b'+1} (-a \rho )  }, \label{post2} 
 \end{equation}
where $b'=b-(\alpha +\alpha')/\alpha \geq 0.$ This suggests that the  heavy-tailed specification  is valid only for $b \geq 2$ as $ (\alpha +\alpha')/\alpha \geq 2.$  Clearly,  $\alpha'= \alpha$ is a straightforward choice.    This formulation achieves  Bayesian  `robustness'  in the sense of   \cite{bdpw16} and the references therein.

The inverse stable prior is worth a more extensive investigation. The following are  specific worthy explorations  for the future:  the  extensive comparison of the inverse stable prior with existing priors (e.g., inverted beta) for rate or inverse scale  (e.g., $\delta$ vs.  $\theta^{-0.5}$)    or inverse variance parameter under different practical settings and/or criteria;  the mitigation of  hypothesis testing procedures and hyper/prior sensitivity analysis;  the investigation of the acceptance rates of  Algorithm 2 for small $\alpha$ values;    the specification of $h(\alpha)$ and application of the hyperprior density in different settings especially concerning predictive distributions;  and the development of  sound inference procedures for more general hierarchical models especially involving the heavy-tailed prior  (\ref{htp}), i.e., with $0<\alpha'<1.$

\section{References}
\renewcommand*{\refname}{}

\section{Appendix A: Proof of Proposition 1}
 
The  posterior distribution kernel   can be written as 
\begin{equation}
   L(\theta | \bm{x}) \cdot IS_{\alpha, \rho } ( \theta) \; \sim \;  e^{-a \cdot \theta}  \;  \theta^b \cdot  \frac{\rho^{1/\alpha}  \; \theta^{-1-1/\alpha}}{\alpha} \;  g_\alpha\left(\; \rho^{1/\alpha}     \theta^{-1/\alpha} \; \right) 
\end{equation}
where $\alpha$ and $\rho$ are the hyperparameters.    To determine the normalizing constant, 

\begin{align}
   \int_{\mathbb{R}^+}   e^{-a \cdot \theta}  \;  \theta^b \cdot \frac{\rho^{1/\alpha}   \; \theta^{-1-1/\alpha}}{\alpha} \;  g_\alpha\left(\; \rho^{1/\alpha}  \;   \theta^{-1/\alpha} \; \right) d \theta  \quad 
&\stackrel{u=\rho^{1/\alpha}    \theta^{-1/\alpha}}{=}\quad  \int_{\mathbb{R}^+}   e^{-a \rho   u^{-\alpha} }  \;  \left( \rho u^{-\alpha}\right)^b  \;  g_\alpha\left( u \right) d u\\
&=\sum_{j=0}^\infty \frac{ (-a \rho)^j }{j!}  \; \rho^b \left( \int_{\mathbb{R}^+}     u^{-\alpha(j+b)}  \;  g_\alpha\left( u \right) d u\right). 
\end{align}

Using the Mellin transform formula for the $\alpha^+$-stable density \citep{zol86},
\begin{align}
\sum_{j=0}^\infty \frac{ (-a \rho)^j }{j!}  \; \rho^b \left( \int_{\mathbb{R}^+}     u^{-\alpha(j+b)}  \;  g_\alpha\left( u \right) d u\right) & = \sum_{j=0}^\infty \frac{ (-a \rho)^j }{j!}  \; \rho^b \left( \frac{\Gamma ( 1 + j+b) }{\Gamma ( 1 + \alpha (j+b))} \right)\\ 
&= \left( \rho^b \;  \Gamma ( b+1)  \right) \; \sum_{j=0}^\infty \frac{ (-a \rho)^j (b+1)_j }{j!\Gamma (\alpha j+ \alpha b +1)}  \\ 
&=  \left( \Gamma ( b+1)\; \rho^{b} \right) \cdot  E_{\alpha,\alpha b + 1}^{b+1} (- a \rho).   \label{res1}  \qed
\end{align}

\section{Appendix B: Test results}

\begin{table}[h!t!b!p!]
\caption{\emph{Means and  MAD of estimates. }} \centerline {
\begin{tabular*}{7.6in}{c|c||c|cc|cc|cc}
 \hline
 \multirow{2}{*}{Data model} & \multirow{2}{*}{$(\alpha, \rho)$} & \multirow{2}{*}{Method}  &  \multicolumn{2}{c|}{$n=15$}  &  \multicolumn{2}{c|}{$n=30$} &  \multicolumn{2}{c}{$n=60$}  \\
& & &   Ave &    MAD&    Ave &   MAD  &   Ave &   MAD  \\
\hline \hline
 \multirow{3}{*}{Poisson$(\theta=4)$} & \multirow{3}{*}{$(0.4, 4)$}  
 &  $\widehat{\theta_1}$ & 3.995& 0.513  & 4.012 & 0.345 & 3.992 & 0.253 \\
 & &  $\widehat{\theta_2}$ & 3.996 & 0.508 &  4.009 & 0.359 &  3.992 & 0.258 \\                                    
& &  MLE & 3.972 & 0.494   &  3.999&  0.346 & 3.986 & 0.247 \\
\hline
\multirow{3}{*}{Rayleigh$\left(\sigma=\sqrt{2/\pi}\right)$} & \multirow{3}{*}{$\left(0.5,  1\right)$}  
     &  $\widehat{\theta_1}$  & 1.322 & 0.303  &  1.303& 0.230 & 1.271 & 0.150 \\
 & &  $\widehat{\theta_2}$  & 1.322 & 0.297 &  1.304 & 0.228  & 1.271 & 0.150 \\                                    
&  &     MLE                                   & 1.332 & 0.330   &  1.301&  0.236& 1.268 &  0.155 \\                   
\hline
\multirow{3}{*}{Half-normal $\left(\sigma=\sqrt{\pi/2}\right)$} & \multirow{3}{*}{$(0.6, \sqrt{2/\pi})$}  
     &  $\widehat{\theta_1}$  & 1.401 & 0.285  &  1.492& 0.261 & 1.540 & 0.242 \\
 & &  $\widehat{\theta_2}$  & 1.401 & 0.289 &  1.493 & 0.265& 1.540 & 0.242 \\                                    
&  &     MLE                                   & 1.815 & 0.606   &  1.684&  0.408& 1.629 & 0.299\\                   
\hline
\multirow{3}{*}{Generalized Exponential $(\theta=2)$}  & \multirow{3}{*}{$(0.8, 1.5)$}  
 &  $\widehat{\theta_1}$ &2.003 & 0.296&  2.031  & 0.276 & 2.026 & 0.220\\
& &  $\widehat{\theta_2}$ &2.003 &  0.303 & 2.032 & 0.267&  2.026 & 0.222  \\                                    
& & MLE & 2.164 & 0.522 &2.084 &  0.367&   2.034&  0.253\\
\hline
\multirow{3}{*}{Exponential $(\theta=1)$} & \multirow{3}{*}{$(0.95,  1)$}  
&  $\widehat{\theta_1}$ & 1.051& 0.066  &  1.054 & 0.073 & 1.046 & 0.079 \\
 & &  $\widehat{\theta_2}$ & 1.051 & 0.065 &  1.054 & 0.072 & 1.046 & 0.081 \\                                    
& &  MLE & 1.061 & 0.241   &  1.035&  0.181 & 1.011 & 0.125\\
\hline
\end{tabular*}
}
  \label{t2}

\end{table}


\vspace{-0.3in}
\begin{thebibliography}{9}
  

\bibitem[Bhadra et al.(2016)]{bdpw16} Bhadra, A., Datta, J., Polson, N. G., Willard, B., 2016. Default Bayesian  analysis with global-local shrinkage priors. Biometrika, 103(4), 955-969.




\bibitem[Beghin and Orsingher(2010)]{bao10}Beghin, L., Orsingher, E., 2010. {\tt Poisson}-type processes governed by fractional and higher-order recursive differential equations. Electronic Journal of Probability, 15(22), 684-709.
 


 
\bibitem[Bingham(1971)]{bing71} Bingham, N.H., 1971. Limit theorems for occupation times of Markov processes. Z. Wahrsch. Verw. Gebiete, 17, 1-22. 

  
\bibitem[Carvalho et al.(2010)]{car10}   Carvalho, C. M., Polson, N. G.,  Scott, J. G., 2010. The horseshoe estimator for sparse signals. Biometrika, 97, 465-480.

\bibitem[Daniels(1999)]{dan99}  Daniels, M.J., 1999. A prior for the variance in hierarchical models,  The Canadian Journal of Statistics, 27(3), 567-578. 

\bibitem[Chambers et al.(1976)]{cms76}  Chambers, J. M.;   Mallows,  C. L., Stuck, B. W., 1976.  A method for simulating stable random variables, Journal of the American Statistical Association,  71(354), 340-344. 
 

\bibitem[Gelman(2006)]{gel06} Gelman, A.,  2006.  Prior distributions for variance parameters in hierarchical models, Bayesian Analysis, 1(3), 515-533. 


\bibitem[Kanter(1975)]{kan75} Kanter, M., 1975.  Stable densities under change of scale and total variation inequalities, The Annals of Probability,  3(4), 697-707.  

  
\bibitem[Kundu(2008)]{kun08}  Kundu, D., 2008. Bayesian inference and reliability sampling plan for Weibull distribution, Technometrics,  50(2), 144 - 154.
 

\bibitem[Iksanov et al.(2017)]{iks17}  Iksanov, A., Kabluchko, Z., Marynych, A., Shevchenko, G., 2017. Fractionally integrated inverse stable subordinators, Stochastic Processes and their Applications, 127(1), 80-106.


\bibitem[Mainardi et al.(2010)]{mmp10} Mainardi, F., Mura, A.,  Pagnini, G.,  2010.  The M-Wright function in time-fractional diffusion processes: a tutorial survey.  Int'l J of Diff'l Equations, vol. (2010).
    
  
\bibitem[Meerschaert et al.(2011)]{mnv11}  Meerschaert, M.M.,   Nane, E.,  Vellaisamy, P., 2011. The fractional {\tt Poisson} process and the inverse stable subordinator. Electronic Journal of Probability, 16, 1600-1620.
  
\bibitem[Meerschaert and Straka(2013)]{mas13}  Meerschaert, M.M.,   Straka, P., 2013. Inverse stable subordinators,  Mathematical Modelling of Natural Phenomena, 8(2), 1-16. 

\bibitem[Miller(1980)]{mil80} Miller, R. B.,  1980,  Bayesian analysis of the two-parameter gamma distribution,  Technometrics, 22(1), 65-69.

\bibitem[Mura et al.(2008)]{mtm08} Mura, A., Taqqu, M.S., Mainardi, F., 2008. Non-Markovian diffusion equations and processes: analysis and  simulations, Physica A, 387, 5033--5064.


\bibitem[Prabhakar(1971)]{pra71} Prabhakar, T.R.,  1971. A singular integral equation with a generalized Mittag-Leffler function in the kernel,Yokohama Mathematical Journal,  19, 7-15.


\bibitem[Piryatinska et al.(2005)]{psw05} Piryatinska, A., Saichev, A.I., Woyczynski, W.A.,  2005.  Models of anomalous diffusion:the subdiffusive case, Physica A: Statistical Physics,  349, 375-424. 
  



  \bibitem[Polson and Scott(2012a)]{pas12a}   Polson,  N.G.,  Scott, J.G., 2012, Local shrinkage rules, L\'evy processes and regularized regression. Journal of the Royal Statistical Society: Series B,  74, 287-311.


  \bibitem[Polson and Scott(2012b)]{pas12b}   Polson,  N.G.,  Scott, J.G., 2012, On the half-Cauchy prior for a global scale parameter, Bayesian Analysis, 7(4), 887-902.

\bibitem[Shukla and Prajapati(1997)]{sap97} Shukla, A.K.,   Prajapati, J.C.,  1997.  On a generalization of Mittag-Leffler function and its properties, J. Math. Anal. Appl., 336, 797-811.




\bibitem[Zolotarev(1986)]{zol86} Zolotarev, V.M. 1986. One-dimensional Stable Distributions: Translations of Mathematical Monographs.  American Mathematical Society, vol 65, Printed in United States of America.
 

\end{thebibliography}
\end{document}